\documentclass[aps,prl,preprint,superscriptaddress]{revtex4}
\usepackage[dvips]{graphicx}

\begin{document}


\title{Tuning mechanical modes and influence of charge screening in nanowire resonators}


\author{Hari~S.~Solanki}
\author{Shamashis~Sengupta}
\author{Sajal~Dhara}
\author{Vibhor~Singh}
\author{Sunil~Patil}
\author{Rohan~Dhall}
\affiliation{Department of Condensed Matter Physics and Materials Science, Tata Institute of Fundamental Research, Homi Bhabha Road,
Mumbai 400005, India}
\author{Jeevak Parpia}
\affiliation{LASSP, Cornell University, Ithaca NY 14853 }
\author{Arnab~Bhattacharya}
\author{Mandar~M.~Deshmukh}
\email[]{deshmukh@tifr.res.in}
\affiliation{Department of Condensed Matter Physics and Materials Science, Tata Institute of Fundamental Research, Homi Bhabha Road,
Mumbai 400005, India}



\begin{abstract}
We probe electro-mechanical properties of InAs nanowire (diameter
$\sim$100~nm) resonators where the suspended nanowire (NW) is also
the active channel of a field effect transistor (FET). We observe
and explain the non-monotonic dispersion of the resonant frequency
with DC gate voltage ($V_g^{DC}$). The effect of electronic
screening on the properties of the resonator can be seen in the
amplitude. We observe the mixing of mechanical modes with
$V_g^{DC}$. We also experimentally probe and quantitatively explain
the hysteretic non-linear properties, as a function of $V_g^{DC}$,
of the resonator using the Duffing equation.
\end{abstract}


\maketitle


Nano electro mechanical systems (NEMS) \cite{Ekincireview}, are
being used extensively to study small displacements
\cite{Displacement1}, mass sensing \cite{Mass1a, Mass1b, Mass1c,
Mass4+Mixer1}, spin-torque effect \cite{Spin2}, charge sensing
\cite{Charge1}, Casimir force \cite{CasimirForce} and potential
quantum mechanical devices \cite{QMresonatora, QMresonatorb}. A
variety of NEMS devices, fabricated using carbon nanotubes
\cite{vera_paper, sciencevanderzant, sciencebachtold,
AKHuttelnanoletters, AKHuttelPRL}, graphene
\cite{graphenescottbunch, grapheneresonator, graphenebachtold},
nanowires (NWs) of silicon \cite{yang-roukes} and by micromachining
bulk silicon \cite{monolithicsculpted}, have been used to probe the
underlying physics at nano scale. In this work we study the
electromechanical properties of doubly clamped suspended n-type InAs
NWs.
In our suspended NW FET, the gate electrode serves three purposes:
first, to modify the tension in the NW, second, to actuate the
mechanical motion of the resonator and third, enabling us to
systematically study the coupling of mechanical properties to the
tunable electron density. As we will show, a tunable electron
density leads to a variable screening length of the order of the
nanowire's cross sectional dimensions. Thus the electro-mechanical
properties enter into a mesoscale regime. Such a variable electron
density is not accessible in carbon nanotubes; the screening length cannot be tuned continuously -- relative to the diameter of the carbon nanotube -- as easily. However, the physics
of charge screening in nanoscale capacitors \cite{chuckblack} and
ferroelectric devices \cite{spaldincapacitor} is intimately
connected to that in our NEMS devices. Taken together, these
observations suggest
 that a geometrical interpretation of capacitance is inadequate
at the nano scale. Additionally, the gate allows us to tune the
resonant frequency non-monotonically due to the competition between
the electrostatic force and the mechanical stiffness ($\sim 1$~N/m)
of the nanowire, a feature expected (but heretofore not studied in
detail) for all electrostatically actuated NEMS. In this article, we
demonstrate that the InAs semiconducting nanowire system manifests
this transition from softening to hardening as the gate voltage is
varied. In addition, mixing of the natural mechanical modes as a
function of $V_g^{DC}$ can be understood in terms of the structural
asymmetries in the resonator. In the non-linear regime we study in
detail hysteretic behavior as a function of $V_g^{DC}$ (unlike the
commonly studied response as a function of drive frequency) and we
show that this can be understood by using the Duffing equation
incorporating the effect of gate voltage. The observed hysteretic
response with $V_g^{DC}$ is an alternate knob for tuning the
nonlinear response of our NEMS devices and can be used for charge
detection \cite{RHBlick}. Our work provides further understanding of
the unique characteristics of NEMS devices operating at room
temperature. The observed behavior can provide information on the
nanomechanics of other systems whose electron density, stiffness or
screening length cannot be so readily tuned.

    The InAs NWs used for this work were grown using metal organic
vapor phase epitaxy (MOVPE) \cite{MOVPE1, MOVPE3}. The NWs are
oriented in the $<111>$ direction and are 80 to 120~nm in diameter
with a length of several micrometers. The substrate used for making
the devices is a degenerately doped silicon wafer with 300~nm thick
SiO$_2$. We have fabricated suspended InAs NW devices by sandwiching
the NWs between two layers of electron-beam resist and then using
electron beam lithography to define the electrodes and suspend them
by depositing $\sim$150~nm Cr and $\sim$250~nm Au after development
followed by \emph{in situ} plasma cleaning \cite{MOVPE3}.  The ohmic
contacts also serve as mechanical supports for the NW suspended
$\sim$200~nm above the surface of SiO$_2$. Fig.\ref{fig:figure1}a
shows the SEM image of a resonator device and scheme for actuating
and detecting the motion of the resonator. All the measurements were
done at 300~K and pressure less than 1~mBar.

    To actuate and detect the resonance we use
the device as a heterodyne mixer \cite{Mass4+Mixer1,
vera_paper,Vera_thesis, Mixer2a, Mixer2b, grapheneresonator}. We use
electrostatic interaction between the wire and gate to actuate the
motion in a plane perpendicular to the substrate. We apply a radio
frequency (RF) signal $V_g(\omega)$ and a DC voltage $V_g^{DC}$ at
the gate terminal using bias-tee. Another RF signal $V_{SD}(\omega
+\Delta\omega)$  is applied to the source electrode
(Fig.\ref{fig:figure1}a). The RF signal applied at the gate
$V_g(\omega)$ modulates the gap between NW and substrate at angular
frequency $\omega$, and $V_g^{DC}$ alters the overall tension in the
NW. The amplitude of the current through the NW at the difference
frequency ($\Delta\omega$), also called the mixing current
$I_{mix}(\Delta \omega)$, can be written as \cite{vera_paper}
\begin{eqnarray}
  I_{mix}(\Delta \omega) &=&I_{oscn}(\Delta\omega)+I_{back}(\Delta \omega) \nonumber \\
  &=&\frac{1}{2}\frac{dG}{dq}(\frac{dC_g}{dz}
z(\omega) V_g^{DC} + C_g V_g(\omega))V_{SD}\label{eq:equation1}
\end{eqnarray}

where $G$ is the conductance of the NW, $q$ is the charge induced by
the gate, $C_g$ is the gate capacitance, $z(\omega)$ is the
amplitude of oscillation at the driving frequency $\omega$ and z
axis is perpendicular to the substrate. The term $I_{back}(\Delta
\omega)=\frac{1}{2}\frac{dG}{dq}C_g V_g(\omega)V_{SD}$ is the
background mixing current which is independent of the oscillation of
the NW and
$I_{oscn}(\Delta\omega)=\frac{1}{2}\frac{dG}{dq}\frac{dC_g}{dz}
z(\omega) V_g^{DC} V_{SD}$ depends on the amplitude of oscillation.
Measuring $I_{mix} (\Delta \omega)$ using a lock-in allows us to
monitor the resonance of the NW as the frequency is swept.
Fig.\ref{fig:figure1}b shows $I_{mix}(\Delta \omega)$ as a function
of $\omega$ for $V_g^{DC}$ = $\pm $ 36.4 V. The sharp feature
corresponds to the mechanical resonance of the NW. We address the
asymmetry of the mixing current signal for $\pm V_g^{DC}$ later.
Fig.\ref{fig:figure1}c shows the plot of conductance ($G$) as a
function of $V_g^{DC}$. The variation of $G$ with $V_g^{DC}$ is very
critical for this scheme of heterodyne mixing to work as it controls
the overall amplitude $I_{mix}(\Delta \omega)$. The hysteresis
observed in the measurement of conductance is typical for our
suspended devices and is associated with charge trap states with
dipolar nature on surface of the nanowire \cite{gatehysteresisa,
gatehysteresisb}.

We can connect the resonant frequency of the fundamental mode,
$f_0$, of a doubly clamped beam at zero $V_g^{DC}$, to the material
properties of the beam as $f_0=C_0 \frac{r}{l^2}V_s$ where
$V_s=\sqrt{\frac{E}{\rho}}$ is the velocity of sound; $r$ is the
radius of the beam, $l$  is the length of the beam, $E$ is the
Young's modulus, $\rho$ is the density of the material and
$C_0=1.78$. Fig.\ref{fig:figure1}d shows a plot for $V_s$ that does
not vary much from the bulk value \cite{bulkvalues} (dashed line)
for eight different devices. The scatter around the $V_s$ calculated
using bulk values could be due to the relative volume fraction
contribution of the amorphous layer around the NWs; this needs
further detailed study.

    Fig.\ref{fig:figure2}a and 2b show the colorscale plot of
    $I_{mix}(\Delta \omega)$ as a function of $V_g^{DC}$ and $\omega$ on
a logscale spanning more than 3 decades. Data is taken by sweeping
$V_g^{DC}$ for each value of frequency. In the data from both the
devices we see a symmetric evolution of the resonant frequency as a
function of $V_g^{DC}$. The parabolic behavior is expected as the
attractive force exerted by the gate on the wire is given by
$F_{DC}=\frac{1}{2}(V_g^{DC})^2 \nabla C_g$. An increase in
$V_g^{DC}$ enhances the tension in the NW\cite{sapmaz}. We now
discuss the particular W-shaped dispersion of modes as a function of
$V_g^{DC}$. As  $|V_g^{DC}|$ is increased, initially the mode
disperses negatively and after a certain threshold voltage
$V_g^{th}$ it disperses with a positive slope. Although purely
negatively and
 positively dispersing modes have been studied in detail before
 by Kozinsky \emph{et al.} \cite{kozinsky}, we observe the crossover regime
 where these interactions compete.  This particular dispersion can be
understood using a toy-model in which a wire is suspended from a
spring of force constant K$_i$ above the substrate. The wire and the
substrate make up the two electrodes of a capacitor. There are two
consequences of increasing $|V_g^{DC}|$: first, it changes the
equilibrium position by moving the NW closer to the substrate and
second, it makes the local potential asymmetric and less steep. The
effective force constant, K$_{eff}$, is reduced, resulting in
negative dispersion of the mode for $|V_g^{DC}|<|V_g^{th}|$. If the
intrinsic force constant K$_i$ had been independent of $V_g^{DC}$,
the modes would always disperse negatively when motion occurs
perpendicular to the gate plane. However, in general K$_i$ can be
written as $k+\alpha (V_g^{DC})^2 + \beta (V_g^{DC})^4 +
H.O.(V_g^{DC})$); where $k$, $\alpha$, and $\beta$ are constants.
With $V_g^{DC}\neq0$ then, to a first approximation, K$_{eff}$ $=$
K$_i-\frac{1}{2}(V_g^{DC})^2 \frac{d^2C_g}{dz^2}$. We find that the
W-shape of the dispersion curve can be explained only if one
considers the case where $\beta>0$. The result of such calculations
(Fig.\ref{fig:figure2}c) quantitatively explains the experimental
observations. $V_g^{th}$, the value at which the crossover from
negative to positive dispersion occurs is a function of the
dimensions of the NW and the capacitive geometry of the device. The
effect of the device dimension is clearly seen in a larger value of
$V_g^{th}$ in device-2 shown in Fig.\ref{fig:figure2}b where the
doubly clamped beam is 120~nm thick, as against 100~nm for device-1
in Fig.\ref{fig:figure2}a  (the lengths differ by 200~nm).

Next, we consider another complementary aspect of the data -- the
amplitude of the mixing current. Fig.\ref{fig:figure2}d shows
$I_{mix}(\Delta \omega)$ (log-scale) as a function of $V_g^{DC}$. We
see that the amplitude of the mixing current for the negative values
of $V_g^{DC}$ are significantly larger than those for positive
$V_g^{DC}$ for the same mechanical mode; this is also seen in
Fig.\ref{fig:figure1}b. We now try to understand this asymmetry as
our InAs NW are n-type semiconductors \cite{MOVPE3} as seen in
Fig.\ref{fig:figure1}c. To understand this asymmetry we have carried
out detailed fits of the experimental data for the amplitude of
mixing current as a function of frequency using
Eqn.\ref{eq:equation1}. The amplitude $z(\omega)$ of oscillation at
frequency $\omega$ is given by
\begin{equation}
z(\omega)=\frac{z_{amp}^{reso}}{Q}\frac{\cos(\Delta\phi+\arctan(\frac{\omega_0^2-\omega^2}{\omega
\omega_0/Q}))}{\sqrt{(1-(\frac{\omega}{\omega_0})^2)^2+(\frac{\omega/\omega_0}{Q})^2}},
\label{eq:equation2}
\end{equation}
where $z_{amp}^{reso}$ is the amplitude at resonant frequency
$\omega_0$, $Q$ is the quality factor and $\Delta\phi$ is the
relative phase difference between the $\omega$ and
$\omega+\Delta\omega$ signals that depends on the device parameters
like the contact resistance. Fitting from Eqns.\ref{eq:equation1}
and \ref{eq:equation2} allow us to extract the variation of $Q$ as a
function of $V_g^{DC}$ as shown in Fig.\ref{fig:figure2}e
(\cite{whyQlow}). We have also estimated the amplitude of
oscillation using Eqns.\ref{eq:equation1} and \ref{eq:equation2} by
examining the ratio
$\frac{I_{oscn}(\Delta\omega)}{I_{back}(\Delta\omega)}=\frac{\frac{dC_g}{dz}
z(\omega) V_g^{DC}}{C_g V_g(\omega)}$. A plot of the calculated
amplitude ($z_{amp}^{reso}$) is seen in Fig.\ref{fig:figure2}f. We
see that as the $|V_g^{DC}|$ is increased $Q$ and $z_{amp}^{reso}$
are observed to decrease. We also observe that there are noticeable
differences in $Q$ and $z_{amp}^{reso}$ for positive and negative
values of $V_g^{DC}$. The values of $Q$ and $z_{amp}^{reso}$ are
larger for positive $V_g^{DC}$. One of the possible mechanisms that
can explain this behavior in the amplitude is that with increasing
$V_g^{DC}$ one increases the density of electrons in the NW leading
to reduction in the screening length. This implies that the simple
geometrical capacitance is inaccurate particularly since the
screening length can be comparable to NW diameter at low densities
(at negative $V_g^{DC}$ in our case). In our device
geometry, using Thomas-Fermi approximation \cite{chuckblack},
screening length is around ~20-40 nm (diameter of our devices are
~100 nm) and the distance between the nanowires and gate oxide
~200nm. So, the screening length is a significant fraction of the
diameter and the suspension distance - this plays a critical role in
observing the effect of density gradients within the cross-section
of the nanowire.  In case of single walled carbon nanotube, diameter
is ~1-2 nm and height of suspension is typically 100nm or more
\cite{sciencevanderzant, sciencebachtold, AKHuttelnanoletters,
AKHuttelPRL}. Additionally the screening length in carbon nanotubes
is typically several multiples of the nanotube diameter
\cite{sasaki}, so due to an increased ratio of suspension distance
to diameter and the large screening length compared to the diameter
it is very difficult to observe the physics we discuss for the case
of nanowires in carbon nanotubes. We would like to point out that this is not a
peculiarity of the InAs nanowires and should be seen in other
semiconducting nanowire devices as well with similar dimensions.
Additionally, if one considers the realistic case with non-uniform density of carriers in
the NW due to the device geometry \cite{wucapacitance} the NW will
have a gradient of dielectric constant \cite{inasscreening}. A
gradient of dielectric constant \cite{Kottahausnewnatphys} alongwith
a change in capacitance as a function of the density changes the
capacitive coupling of the NW to the gate. This results in differing
amplitudes for two different electron densities. Our device geometry
with NW diameter comparable to the gap accentuates this effect.

In order to better understand the effect of the gate voltage in
tuning the spatial charge density across the cross-section of the
nanowire we selfconsistently solve three dimensional Poisson's
equation using finite element method for our device geometry. We use
approach followed by Khanal \emph{et al.} \cite{wucapacitance} by
solving $ \nabla \cdot \epsilon_d \nabla \Phi(x,y,z) = \rho(x,y,z)$,
throughout the space of the nanowire and its dielectric environment
(here $\epsilon_d$ is dielectric constant, $\Phi$ is the local
electrostatic potential in the system due to applied gate voltage,
and $\rho$ is space charge density inside the NW). The geometry
consists of a 100~nm diameter and 1.5 $\mu$m long InAs nanowire
clamped by metallic electrodes. The wire is suspended 100~nm above a
300~nm thick silicon-oxide dielectric on the gate electrode. Inside
the nanowire the $\rho(x,y,z)=e(N_d-n(\Phi)+p(\Phi))$ where $e$ is
the charge of an electron, $N_d$ is the density of the n-type
dopants ($\sim 10^{16}$cm$^{-3}$), $n(\Phi)$ and $p(\Phi)$ are the
densities of electrons and holes. The unintentional source of n-type
dopants in our growth is Si and C, from the metal organic precursors
and are assumed to be uniformly distributed throughout the nanowire.
A selfconsistent calculation gives us the distribution of potential
throughout the space and space charge density in the nanowire.
Fig.\ref{fig:figure3}a and 3b show the colourscale distribution of
potential when the $V_g^{DC}=-5$~V. In order to model the
consequences of modifying gate voltage we calculated the
distribution of space charge density through the nanowire for
$V_g^{DC}=-25, -5$ and $5$V. Fig.\ref{fig:figure3}c, 3d, 3e, and 3f
show the result of such a calculation for varying $V_g^{DC}$ for
space charge density in the center of the NW. For positive voltages
the Fermi energy is very close to the conduction band of the InAs
and as a result the charge density is very uniform while behaving as
a metal like system. This simple modeling supports our arguments
that a gradient of electron density can modify the capacitive
coupling and the resulting amplitude. Further analysis is needed to
solve self consistent solutions to the Poisson's equation where the
dielectric constant is itself a function of the density and the
quantitative variation of $Q$ and amplitude with sign of $V_g^{DC}$.
Our measurements suggest a way to tune the efficiency of actuation
by tuning the density of carriers.

    We next consider three other features of our data -- first, the
presence of other mechanical modes near the fundamental mode;
second, the mixing of modes as a function of the  $V_g^{DC}$, and
third, the non-linear properties of NW oscillators driven to large
amplitudes. Fig.\ref{fig:figure4}a and 4b show the plot of
$I_{mix}(\Delta \omega)$ as a function of $V_g^{DC}$ and $\omega$
for device-3 and device-2. It is well known that for a doubly
clamped beam with no tension, $ f_n=C_n
\frac{r}{l^2}\sqrt{\frac{E}{\rho}}$ with $C_0=1.78$, $C_1=4.90$ and
$C_2=9.63$ for the transverse modes. It is clear that if the
fundamental mode ($f_0$) is described by a mode with zero nodes and
moving in a plane perpendicular to the substrate, the other observed
modes, in the frequency range near $f_0$ (Fig.\ref{fig:figure4}a,
4b), cannot be defined by $f_1$ and $f_2$. We have observed the
anticipated $f_1$ and $f_2$ modes at higher frequencies. The other
modes beside the fundamental in Fig.\ref{fig:figure4}a, 4b, are
explained due to geometrical asymmetry along the diameter in the NW
(Fig.\ref{fig:figure4}c, details of this calculation provided in
Supporting Information). These are the modes involving motion in a
plane that is not perpendicular to the substrate. This would explain
the less steep slope of the dispersion as a function of $V_g^{DC}$.

    Fig.\ref{fig:figure4}a and 4b also show the mixing of the modes as a function of $V_g^{DC}$. The
mode mixing can be seen clearly in {Fig.\ref{fig:figure4}d which
shows a close-up of the data in Fig.\ref{fig:figure4}b. Displacement
along the transverse direction $y$ (perpendicular to $z$) will
weakly affect the capacitance because of any slight asymmetry in the
physical structure of the NW. The coupling coefficient $\frac{1}{2}
\frac{\partial^2C}{\partial z\partial y}(V_g^{DC})^2$ appearing in
the potential energy gives rise to mode-mixing (see Supporting
Information). For the device shown in {Fig.\ref{fig:figure4}b the
minimum frequency gap in the region of level repulsion is $0.25 \pm
0.02$~MHz. The asymmetry in the amplitude of the modes away from the
region of mixing can also be understood within this model.

We  next consider the non-linear response of these NEMS oscillators.
Due to electrostatic actuation, the potential energy of the
oscillator is asymmetric about the equilibrium position and has the
form $V(z)= \frac{1}{2} K_{eff} z^2 + \theta z^3 + \mu z^4 +
H.O.(z)$, where $K_{eff}$, $\theta$ and $\mu$ are coefficients
depending on $V_g^{DC}$. We have experimentally probed the
non-linear and hysteretic response of the device.
Fig.\ref{fig:figure5}a, 5b, 5c, and 5d, show the experimentally
measured non-linear response of this device leading to hysteretic
behavior as function of $V_g^{DC}$ with increasing amplitude of
driving force. Here, $V_g^{DC}$ is swept at a given drive frequency
to measure the mixing current and several such $V_g^{DC}$ sweeps are
collated (the distinct response of sweeping the frequency at a fixed
$V_g^{DC}$ -- a common mode to study non-linear response -- is
described in supporting information). There are two features that we
would like to point out -- first, in the region at the bottom of the
W-shaped dispersion curve, two branches of the same mode merge into
one broad peak where the oscillator has large amplitude over a wide
range of $V_g^{DC}$ and second, whenever during the $V_g^{DC}$
sweep, one crosses the region with a local negative value for
$\frac{df_0}{dV_g^{DC}}$ (here, $f_0$ is the resonant frequency at a
particular $V_g^{DC}$), shows a curved hysteretic \emph{j}-shaped
response (indicated by $\star$), seen in Fig.\ref{fig:figure5}b. In
order to understand and explain the experimentally observed
hysteretic response as a function of $V_g^{DC}$, we have used the
Duffing equation \cite{landau} for our resonator. The result of such
a calculation for amplitude is seen in Fig.\ref{fig:figure5}e, 5f,
5g, and 5h, with increasing amplitude of driving force. To calculate
the amplitude we have only used the observed dispersion relation (W
shape) as input for Duffing equation. There is a qualitative
agreement between the experimentally measured data shown in
Fig.\ref{fig:figure5}a-d and results of calculation using the
Duffing equation shown in Fig.\ref{fig:figure5}e-h. We find that for
every increase in excitation amplitude by 100~mV corresponds to an
increase of a factor of 2.5 in the anharmonic component of the
Duffing equation (from observing the calculated data). Additional
aspect of the nonlinearity of the oscillator is also seen in the
evolution of dispersion near $V_g^{DC}=0$ as the actuation amplitude
is gradually increased from 100~mV to 400~mV in
Fig.\ref{fig:figure5}a-d. The negative dispersion is due to the
softening of electrostatic force and with larger amplitude of
oscillation the effective spring constant changes, as the wire
samples a region with varying electric field; this difference is
clearly seen in the dispersion near $V_g^{DC}=0$ for the data shown
in Fig.\ref{fig:figure5} a and d.

Fig.\ref{fig:figure5}i, shows
the line plot (along the dashed line in Fig.\ref{fig:figure5}f) of
calculated amplitude for different sweep direction of $V_g^{DC}$ and
the resulting hysteresis.  The observed hysteretic
response, as a function of $V_g^{DC}$, is quite different from the hysteretic response as a function of
drive frequency ( discussed in supporting information ). This non-linear response of our devices
with $V_g^{DC}$ may be utilized for charge detection \cite{RHBlick} as near the onset of non-linearity the
change in amplitude as a function of $V_g^{DC}$ is very large.

In summary, we have studied the electromechanical properties of
doubly clamped InAs NW resonators. Their size and tunable electron
density allow us to map behavior that has not been manifested in a
single device. We have observed and quantitatively explained the
competition between the softening of stiffness of the restoring
force of the resonator due to the variation of the electrostatic
force under variable gate voltage. At larger voltages, the
stretching of the nanowire leads to increased stiffness resulting in
a non-monotonic dispersion of the fundamental mode with $V_g^{DC}$.
The screening of electric fields due to the variation in the density
of the electrons in our suspended FET devices modifies the amplitude
because the variation of the screening length spans the cross
sectional dimension of our nanowire. Further, the non-linear
properties of our device can be understood qualitatively using the
Duffing equation that explains the hysteretic response of the
amplitude as a function of gate voltage. Thus in a single device, we
demonstrate, separate and account for three diverse behaviors. Our
measurements indicate that measuring electromechanical response
influenced by charge screening could lead, in the future, to new
ways to probe spin physics by exploiting spin-dependent charge
screening \cite{spindependentscreening}. Probing the physics by
tuning electron density in NEMS devices may help probe the role of
defects \cite{defectsscreening} and electron hopping as one moves
from insulating to conducting regimes. Control over the non-linear
dynamics may be achievable by controlling mode mixing and DC gate
voltage.

This work was supported by Government of India. J. Parpia was supported by DMR-0457533.


\begin{thebibliography}{42}
\bibitem[{why()}]{whyQlow}
\bibinfo{title}{Our devices show relatively low $Q$ ($\sim$
150-600 ) as
  gold is soft and clamping losses could result}.
\end{thebibliography}

\newpage

\begin{figure}
\includegraphics[width=120mm]{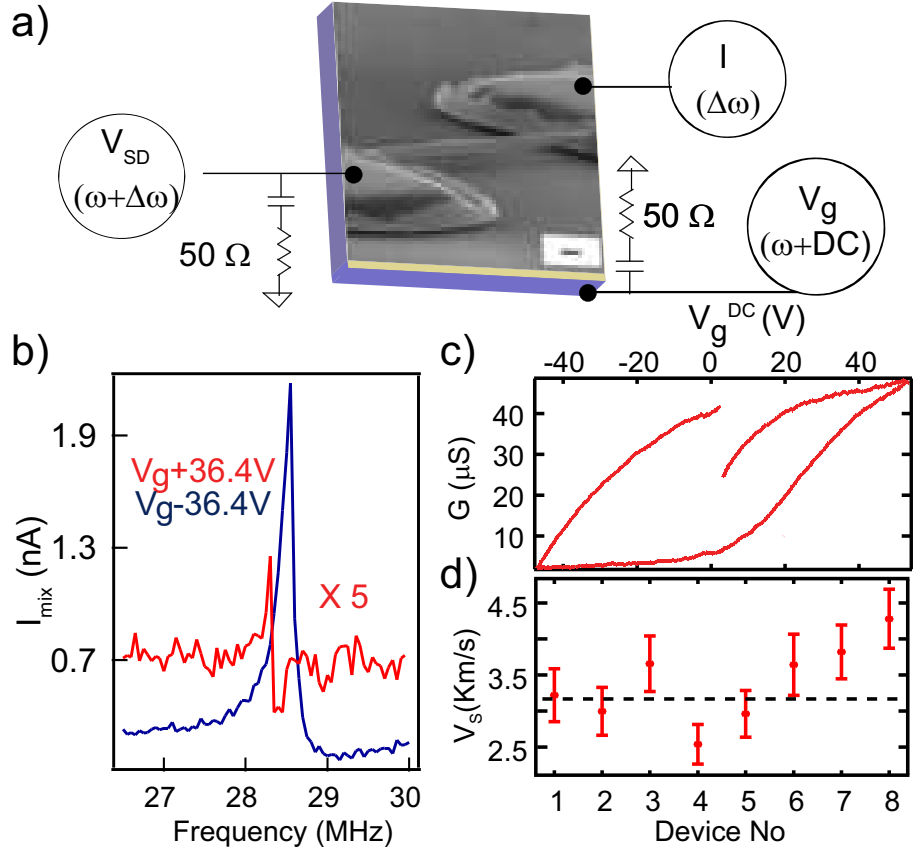}
\caption{\label{fig:figure1} (color online) a) Tilted angle SEM
image with the circuit used for actuation and detection of resonance
for an InAs NW resonator. The diameter of the wire is 100~nm and the
length of NW is 2.9~$\mu$m. The scale bar indicates a length of
200~nm. b) The mixing current ($\Delta \omega/2\pi$ = 17 KHz) as a
function of frequency for two values of $V_g^{DC}$ (mixing current
shown for +ve $V_g^{DC}$, is 5 times of its original value). c)
Variation of the conductance as a function of DC gate voltage. d)
The plot of sound velocity $V_s=\frac{f_0l^2}{1.78r}$ calculated
using the measured frequency ($\omega/2\pi$) of the fundamental mode
of the InAs NW resonators and geometrical values. The dashed line
indicates the speed of sound obtained for $V_s=\sqrt{E /\rho}$ using
bulk values for $E$ and $\rho$ .\cite{bulkvalues}}
\end{figure}

\newpage

\begin{figure}
\includegraphics[width=125mm]{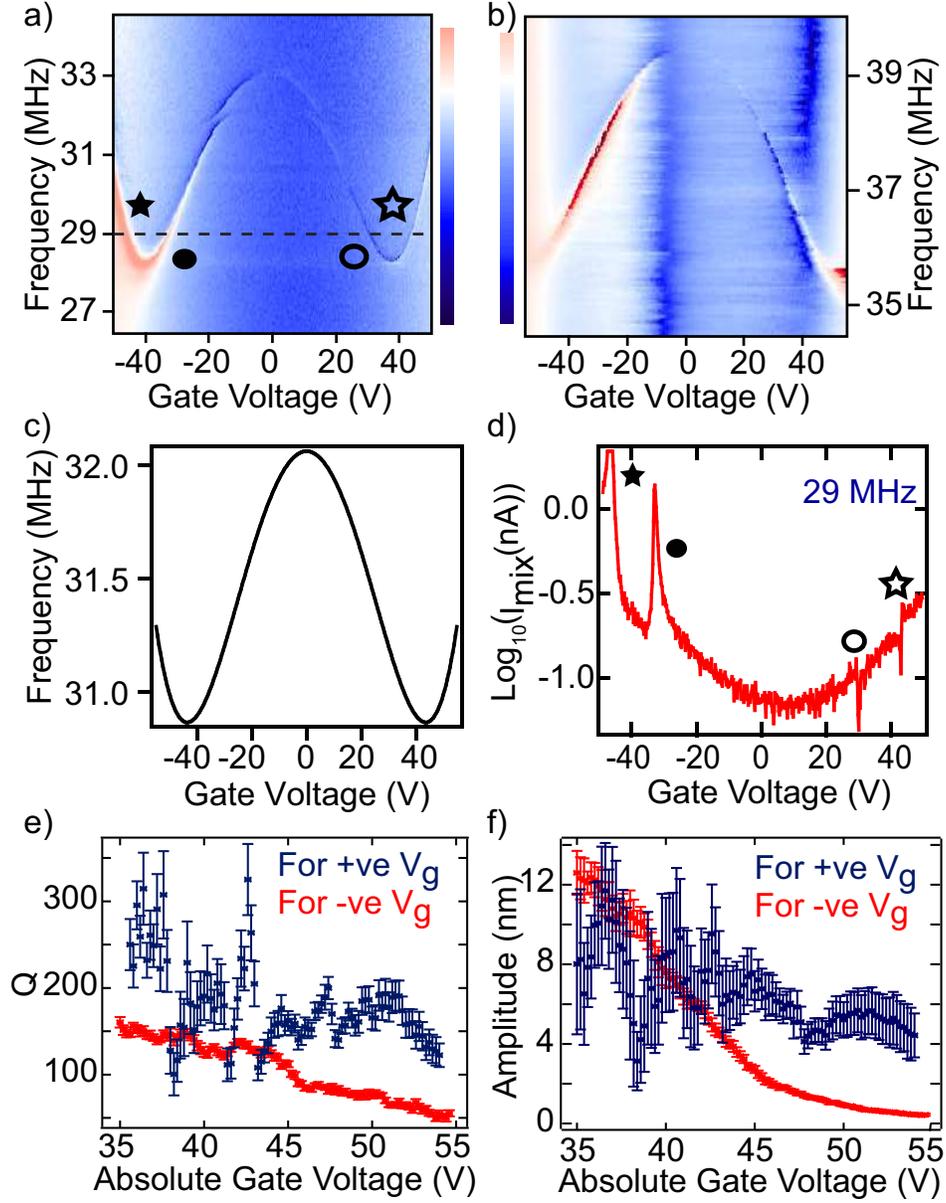}
\caption{\label{fig:figure2} (color online) a) and b) Color logscale
plots of mixing current as a function of $V_g^{DC}$ and
$\omega/2\pi$ for two devices (for device-1,
~Fig.\ref{fig:figure2}a, diameter d=100 nm, length l=2.9 $\mu$m, for
device-2, ~Fig.\ref{fig:figure2}b, d=120 nm, l= 3.1 $\mu$m). c)
\emph{Calculated} dispersion as a function of $V_g^{DC}$. d)
Lineplot at 29 MHz for device-1 (dashed line  in
~Fig.\ref{fig:figure2}a). e) The plot of the Q as a function of
$V_g^{DC}$ for device-1. f) The plot of the ``amplitude"
$z_{amp}^{reso}$ as a function of $V_g^{DC}$ for device-1. (The red
and blue traces in Fig.\ref{fig:figure2}e, and in
~Fig.\ref{fig:figure2}f, show the data for negative and positive
gate voltages).}
\end{figure}

\newpage

\begin{figure}
\includegraphics[width=125mm]{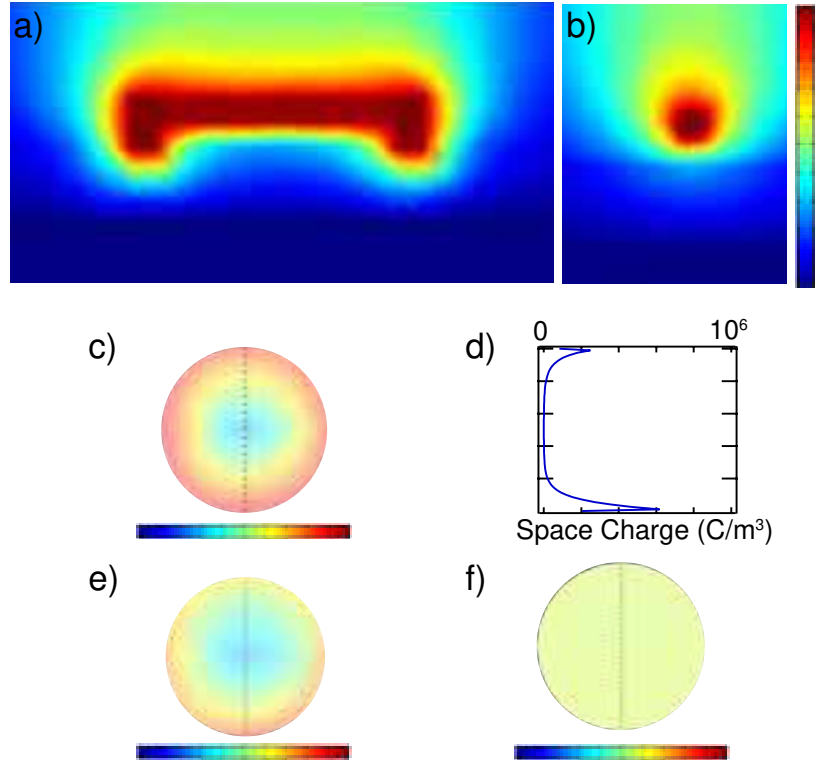}
\caption{\label{fig:figure3} (color online) a) and b) Shows  the
result of a FEM based self-consistent solution of Poisson's equation
giving the potential around a doubly clamped suspended nanowire
device 100~nm in diameter and 1.5~$\mu$m~ long. The separation of
the NW and 300~nm thick SiO$_2$ is 100~nm. The back plane of the
nanowire is the gate held at -5V and the two terminals of the wire
are grounded. The maximum of the colorscale bar (red)is 0V and the
minimum (blue) is -5~V. c) The space distribution in the cross
section of the nanowire for $V_g^{DC}$ = -25V shows the gradient.
The log-colorscale below varies from 10 C/m$^3$ (blue) to $10^6$
C/m$^3$ (red). d) Plot of space charge density in the vertical
direction at $V_g^{DC}$ = -25V through the middle of the wire. The
asymmetry along the vertical direction due the gate below the wire
is clearly seen. e) The space distribution in the cross section of
the nanowire for $V_g^{DC}$ = -5V shows the gradient. The
log-colorscale below varies from 10$^3$ C/m$^3$ (blue) to $10^5$
C/m$^3$ (red). f) The space distribution in the cross section of the
nanowire  for $V_g$= 5V shows uniform space charge distribution as
the electron density in the nanowire is increased when the nanowire
field effect transistor is turned on. The log-colorscale below
varies from 10$^3$ C/m$^3$ (blue) to $10^5$ C/m$^3$ (red).}
\end{figure}

\newpage

\begin{figure}
\includegraphics[width=125mm]{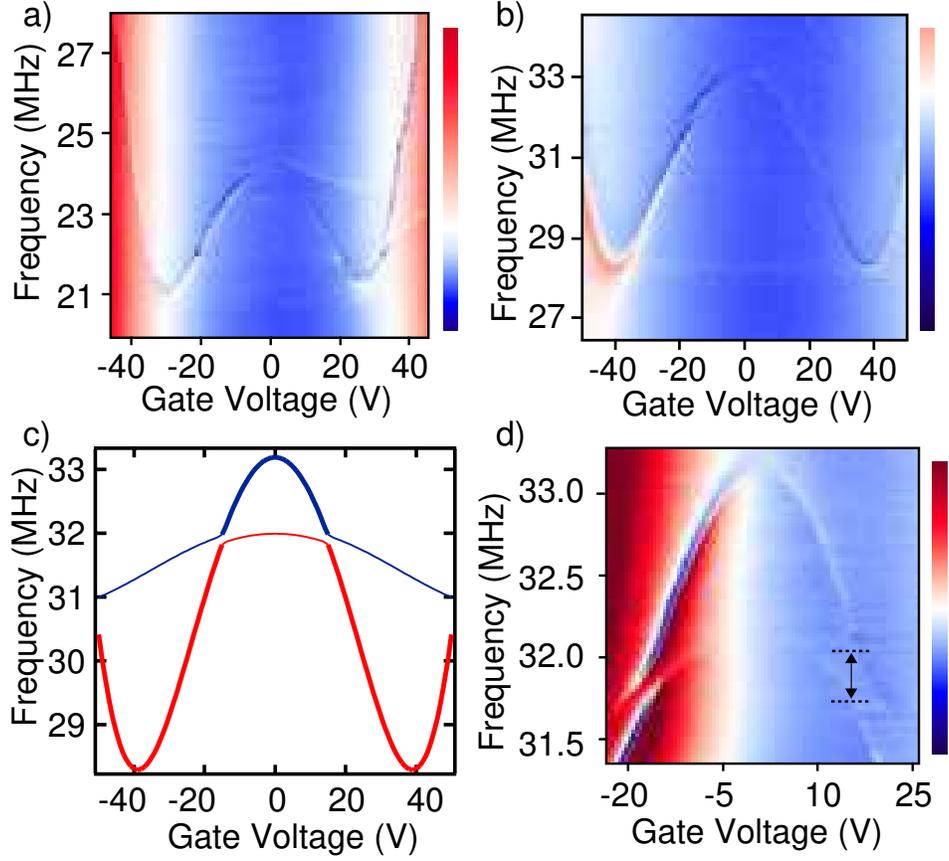}
\caption{\label{fig:figure4} (color online) a) and b) Shows mode
mixing for device-3 and device-1 respectively (for device-3, d=103
nm and l=3.1 $\mu$m). c) \emph{Calculated} dispersion for mode
mixing using asymmetry in the wire. d) Zoomed-in view for device-1
(Fig.\ref{fig:figure4}b) which shows the mixing of the modes as a
function of $V_g^{DC}$. Minimum separation between the modes,
indicated by the double-headed arrow, is $0.25\pm0.02$ MHz.}
\end{figure}

\newpage

\begin{figure}
\includegraphics[width=80mm]{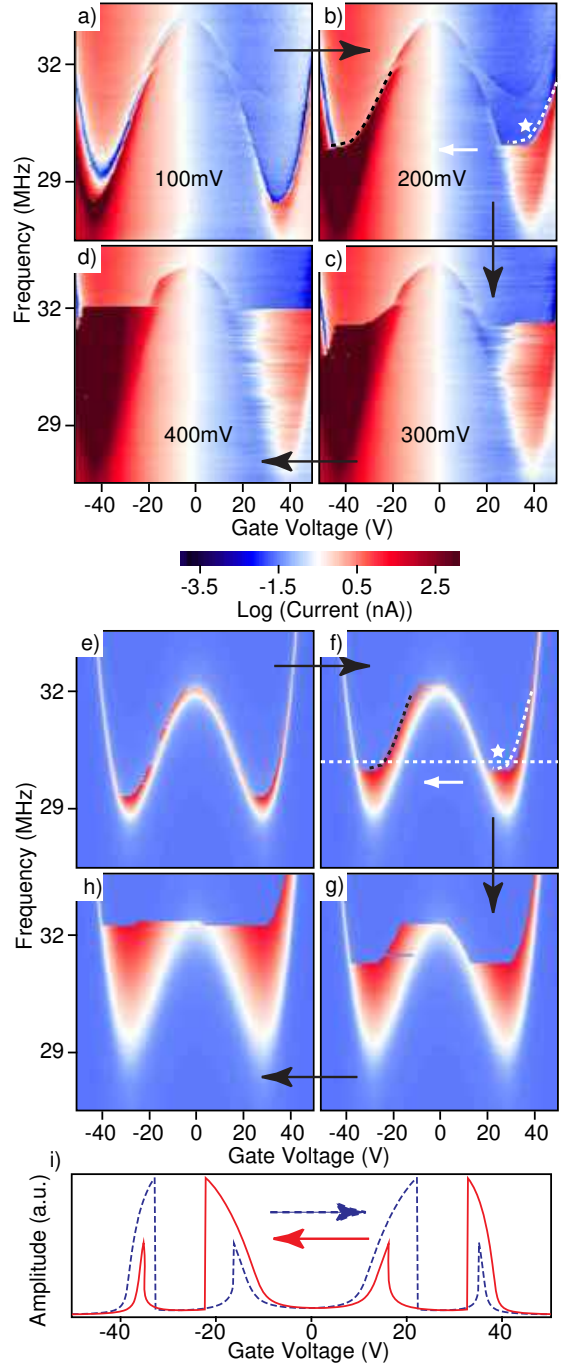}
\caption{\label{fig:figure5} (color online) a), b), c), and d) Color
logscale plots of mixing current from device-1 as a function of
$V_g^{DC}$ and $\omega$ in non-linear regime with increasing
amplitude of driving force starting from 100 mV to 400 mV
respectively. e), f), g), and h) Colorscale plots of
\emph{calculated} amplitude of a resonator in the non-linear regime
using Duffing equation with increasing amplitude of driving force
from Fig.\ref{fig:figure5}e to Fig.\ref{fig:figure5}h. i) Shows the
amplitude for the two directions of $V_g^{DC}$ which indicates
hysteresis at constant frequency 31 MHz (red and blue arrows show
corresponding gate sweep direction).}
\end{figure}

\end{document}